\documentclass[aps,prb,twocolumn,superscriptaddress]{revtex4}
\newcommand{\bvec}[1]{\mbox{\boldmath$#1$}}

\usepackage{graphicx}
\begin{document}

%\preprint{}

\title{Superconductivity due to spin fluctuations originating from 
multiple Fermi surfaces in a 
double chain superconductor Pr$_2$Ba$_4$Cu$_7$O$_{15-\delta}$}


\author{Tsuguhito Nakano}
\author{Kazuhiko Kuroki}
%\email[]{tnakano@vivace.e-one.uec.ac.jp}
%\thanks{}
\affiliation{Department of Applied Physics and Chemistry, The
University of Electro-Communications, Chofu, Tokyo 182-8585, Japan}
\author{Seiichiro Onari}
\affiliation{Department of Applied Physics, The University of Nagoya,
Nagoya 464-8603, Japan}


\date{\today}

\begin{abstract}
The mechanism of superconductivity in Pr$_2$Ba$_4$Cu$_7$O$_{15-\delta}$ 
is studied using a quasi-one dimensional double chain
 model with appopriate hopping integrals, on-site $U$, 
and off-site repulsion $V_1$. Applying the
 fluctuation exchange method to this model and solving the Eliashberg equation,
 we obtain the doping dependence of superconductivity that is consistent 
with the experiments. The superconducting gap has an extended $s$-wave-like  
form, which gives a temperature dependence of the spin lattice relaxation rate 
that does not contradict with the experimental results.
\end{abstract}
%\keywords{}

\maketitle


It goes without saying that unconventional superconductivity (SC) appears on
the CuO$_2$ planes or ladder structures in some
cuprates.\cite{BM,Uehara} Recently, SC in the cuprates has been found 
in another structure, namely, the CuO double chain. 
Pr$_2$Ba$_4$Cu$_7$O$_{15-\delta}$  (Pr247/$\delta$), which consists of
metallic CuO double-chain and semiconducting single-chain besides the
Mott insulating CuO$_2$ plane, shows SC with $T_c^{max} \sim 15$ K in
a moderate oxygen defect concentration range of $\delta =0.2-0.6$  
\cite{FR,Matsukawa,YYamada},
which controls the band filling of the double chain block. 
%In Pr-substituted cuprates, the SC in CuO$_2$ plane is
%strongly suppressed since the doped carriers are trapped in the
%Pr$4f$-O$2p_\pi$ hybridized orbital.\cite{FR} In
%PrBa$_2$Cu$_3$O$_{7-\delta}$ (Pr123), which has a single-chain and
%CuO$_2$ planes, the single-chain shows the semi-conducting behavior
%although the carrier concentration can be controlled. \cite{Peng,
%Grevin} On the other hands, PrBa$_2$Cu$_4$O$_8$, which has double-chains
%and planes, shows the metallic resistivity but, in turn, the changing
%the amount of carrier is not allowed. \cite{Horii} Pr247 resides both
%features: namely it has ``carrier-number-controllable'' double-chain
%system.
Recent NQR study has revealed that the SC occurs in the 
double-chain, and further observed a ``charge freezing'' just like the one 
observed in PrBa$_2$Cu$_4$O$_8$ (Pr124)\cite{Watanabe, Sasaki, Fujiyama}
at, and only at, $\delta=0$, 
which implies that the double chain is near $1/4$-filling in Pr247$(\delta=0)$ 
(although there remains some ambiguity in the band filling estimation). 
\cite{Mizokawa,Takenaka, Seo} 

%and the situation of double chain of Pr247/$\delta=0$ and Pr247(super) are
%smiler to that of Pr124 and YBa$_2$Cu$_4$O$_8$ (Y124)
%respectively. \cite{Watanabe, Sasaki,Fujiyama}
%Especially in Pr247/$\delta=0$ and Pr124, the local charge ordering
%called ``charge freezing'' are observed, which implies the off-site
%interaction may play important role because of the system being near the
%$1/4$ filling. \cite{Mizokawa, Takenaka, Seo}
%Since the resistivility and $1/T_1$ of Pr247/$\delta=0$ are very similar
%to those of Pr124, the superconductivity is thought to occure in carrier
%doped double chain. 

Theoretically, Sano {\it et al}. proposed a superconducting mechanism 
using a one dimensional (1D) double chain model,\cite{SO} 
which is essentially 
based on a superconducting mechanism proposed generally by Fabrizio
\cite{Fabrizio} and also studied by one of the present authors\cite{Kuroki} 
about a decade ago.
In a purely 1D double chain system, 
the band dispersion can have a double well structure,
resulting in four Fermi points for appropriate band fillings, which 
in turn results in an opening of the spin gap and dominating 
superconducting correlation.
Sano {\it et al.} 
determined the tight binding (TB) parameters for a purely 1D system 
using the LDA band dispersion between the X and S point for the double
chain of YBa$_2$O$_4$O$_8$ (Y124) (See Fig. \ref{fig2})\cite{Ambrosch}.
From this band structure, they have suggested that the appearance of SC
upon oxygen reduction is due to the increase of the number of Fermi
points from two to four as the band filling is increased from $\sim
1/4$-filling.

Although we believe that this theory is correct in that 
the existence of four Fermi surfaces (FS) is essential in the occurrence of 
SC, there remain several problems, which have motivated the present study.
First, if we look into the band structure in the entire 
two dimensional Brillouin zone (BZ) shown in Fig.\ref{fig2},
the number of the FS along the $\Gamma$-Y line remains to 
be four down to very low band fillings 
(i.e., the inner FS is 2D),\cite{Yu,comment} 
so the variance of the FS around $1/4$-filling 
is not simply a change of their number between 
two and four as in the pure 1D theory. Therefore, the reason for the 
absence of $T_c$ before oxygen reduction is not clear. 
Secondly, a previous estimation of the spin gap using density matrix 
renormalization group (DMRG) has shown that the spin gap in a double chain 
Hubbard model is, if any, very small at $1/4$-filling,\cite{Arita} 
although the values of the 
hopping integrals taken there do not directly correspond 
to Pr247. Moreover, a recent DMRG analysis shows that 
the spin gap, if any, is too small to be estimated numerically when 
the ratio between the interchain and the 
intrachain hopping is smaller than $\sim 0.3$,\cite{Okunishi} 
while this ratio is estimated to be around $0.2$ in Pr247
from first principles calculations. \cite{comment}
% have to say that Pr247 and Y248 have essentially the same band
Thirdly, it is of interest whether the relatively high $T_c$ of 15K 
can be explained within a Hubbard-type model 
with the above mentioned ratio as small as $\sim 0.2$,
because it is obvious that SC does not 
occur when this ratio is too small, namely, when the system is
essentially a single chain repulsive Hubbard model.\cite{Solyom} 
%%However, if we focus on the band dispersion from the 
%%point, it is likely that the change in the number of Fermi points does
%%not occur unless at very low filling.
%This fact tellus that there may be slightly two dimensionality in the band
%dispersion and it is significant to study this material with the Fermi
%liquid approach.  
To resolve these problems, 
%investigate the another possibility mechanism of the
%oxygen reduction induced superconductivity, 
we propose in this paper 
a spin fluctuation mediated pairing mechanism, where the 
origin of the spin fluctuation is indeed the presence of 
multiple FS.
We adopt the fluctuation exchange
(FLEX) method to a quasi-one dimensional (Q1D) extended
Hubbard model for the CuO double chain.\cite{Bickers,Esirgen} A set of TB
parameters are determined from the results of LDA calculation for
Y124. \cite{Yu,comment} 
By solving the linearized Eliashberg equation, we obtain a 
finite $T_c$ for an extended $s$-wave SC, 
whose band filling dependence is qualitatively
similar to the experimental results.

%In fact, although there is no first principles calculation study for Pr247,
%the band dispersion of double chain may be hardly differ from that of Y124. \cite{Oguchi}
%A set of hopping parameters is determined so that the tight binding
%dispersion with these parameters can well reproduce the LDA calculation
%results for the CuO chains both in the $k_a=0$ and $k_a=\pi$
%directions. 
% By introducing the Coulomb interaction between the nearest neighboor
%(n.n.) sites, the superconductivity turns out to vanish around $1/4$-filling. 

\begin{figure}
\begin{center}
 \includegraphics[width=7.4cm]{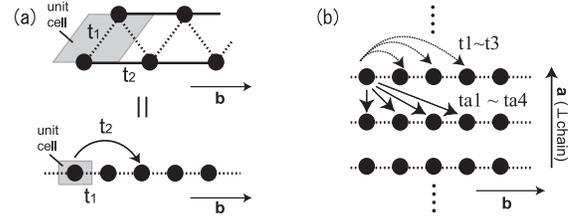}
\end{center}
\caption{\label{fig1} (a)The effective single band 1D double chain
 model.(b) Q1D double chain model. }
\end{figure}

\begin{figure}
\begin{center}
 \includegraphics[width=7.4cm]{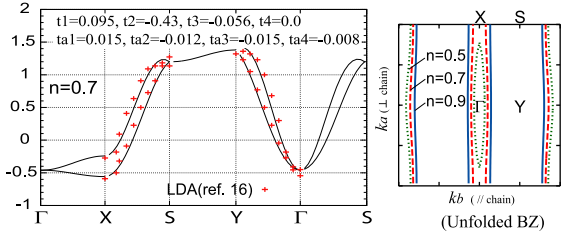}
\end{center}
\caption{\label{fig2} (Color online)(a) The TB band dispersion for
$n=0.7$ with the determined parameters along with the LDA results from
 ref. \onlinecite{Yu}. (b) The FS (for the noninteracting case) for several band fillings.}
\end{figure}

%Since there is the sufficient difference of the energy level between
% the Cu $3d$ and O $2p$ orbital, 
Note that the original model on the zigzag lattice (which has two sites
in a unit cell and thus results in a two band system) can be mapped to a
single band model having distant hopping integrals as shown
Fig. \ref{fig1}. Such a mapping is often adopted in the study of zigzag
chains\cite{Fabrizio,Seo}. Note that after this single band mapping, the
BZ will be unfolded in the $k_b$ direction. Hereafter, our FLEX results
will be shown in the unfolded BZ. 
%, so that one can obtain the effective single band model which presents CuO
%double chain essentially. %Moreover, the zig-zag chain can be mapped to
%straight aligned chain %model with same values of hopping parameters (Fig. {\ref{@}}). 
The Hamiltonian of the extended Hubbard model for the double chain is given by 
\begin{equation}
 H=\sum_{i,j,\sigma}t_{ij}c_{i\sigma}^{\dagger}c_{j\sigma} + U \sum_{i}n_{i\uparrow}n_{i\downarrow}+\frac{1}{2}\sum_{i,j}V_{ij}n_{i}n_{j},
\end{equation}
%where i and j denote the site, $c_{i\sigma}^{\dagger}$ ($c_{i\sigma}$)
%is the creation (annhilation) oparator of Fermion
where $c_{i\sigma}^{\dagger}$ creates an electron of spin $\sigma$ at
site $i$, $t_{ij}$ is a hopping parameter between site $i$ and $j$, and
$U$ and $V_{ij}$ is an electron-electron interaction between such
sites. The values of $t_{ij}$ are determined by fitting the TB
dispersion to the results of the LDA calculations for the double chains of Y124.\cite{comment}
%As mentioned above, we determine the values of hopping parameters so that
%the TB band dispersion with these parameters can reproduce the energy
%band of double chain of Y124 obtained by LDA. 
The fitting result is shown in Fig. \ref{fig2}. With the obtained values
of the hopping parameters, we have found that for the noninteracting
case the inner FS around the $\Gamma$ point remains unless $n \leq 0.1$
($n$=band filling=number of electrons/number of sites), although a
topological change does occur near the X point at $n \sim 0.3$
(Fig. \ref{fig2}(b)), which corresponds to the change of the number of
Fermi points discussed in ref. \onlinecite{SO}.  
In the FLEX, the dressed Green's function is obtained by solving the
Dyson's equation $G(k)=[G_0^{-1}(k)-\Sigma(k)]^{-1}$ self-consistently,
where $G_0(k)$ is the undressed Green's function and $\Sigma(k)$ is the
self energy written by the spin and charge susceptibilities
$\bar{\chi}_{s,c}(q)=\bar{\chi}(q)[I+\bar{V}_{m,d}(q)\bar{\chi}(q)]^{-1}$
with $k \equiv (\bvec{k},\varepsilon_n=(2n+1)\pi T)$ and $q \equiv
(\bvec{q},\omega_l=2l\pi T)$. $\bar{\chi}(q)$ is the irreducible
susceptibility which is composed of a product of $G(k)$, and
$V_{m,d}(q)$ is the magnetic and density coupling vertices described
with $U$ and $V_{ij}$. $\bar{\chi}(q)$, $\bar{\chi}_{s,c}(q)$ and
$V_{d,m}(q)$ are the matrices indexed by the initial (final) relative
displacement of particle-hole pair $\Delta \bvec{r}$ ($\Delta
\bvec{r}'$). The sizes of these matrices increase as the number of
off-site interactions increases; namely if $U$ and the nearest neighbor (n.n.)
repulsion $V_1$ exist, the matrix size becomes $3 \times 3$. The pairing
interactions are given by $\Gamma_s(k,k') \propto
\sum_{\Delta\bvec{r},\Delta\bvec{r'}}[\frac{3}{2}\bar{V}_m\bar{\chi}_s\bar{V}_m-\frac{1}{2}\bar{V}_d\bar{\chi}_c\bar{V}_d](k-k')$
for singlet pairing and $\Gamma_t(k,k') \propto \sum_{\Delta\bvec{r},\Delta\bvec{r'}}[-\frac{1}{2}\bar{V}_m\bar{\chi}_s\bar{V}_m-\frac{1}{2}\bar{V}_d\bar{\chi}_c\bar{V}_d](k-k')$
for triplet pairing. With the obtained $G(k)$ and $\Gamma(k,k')$, the
linearized Eliashberg equation
$\lambda\Delta(k)=-\frac{T}{N}\sum_{k'}\Gamma(k-k')G(k')G(-k')\Delta(k')$
is solved. $T_c$ is defined as $T_c \equiv T(\lambda=1)$. In order to
assure the convergence of the calculation, the system size and the number of
Matsubara frequencies are taken as $N=512 \times 16$ sites and 16384,
respectively, when only $U$ exists, and $256 \times 16$ sites and 16384
are taken when both $U$ and $V_1$ are present. As for the band filling
range, we have investigated in the range of $n=0.3$ to near
half filling. In the following, we show the irreducible susceptibility
$\chi(\bvec{q})$, spin and charge susceptibilities
$\chi_{s,c}(\bvec{q})$ defined as the largest eigenvalues of
$\bar{\chi}(\bvec{q})$, $\bar{\chi}_{s,c}(\bvec{q})$. These along with
the gap function are plotted at the lowest Matsubara frequency.

\begin{figure}
\begin{center}
 \includegraphics[width=6.4cm]{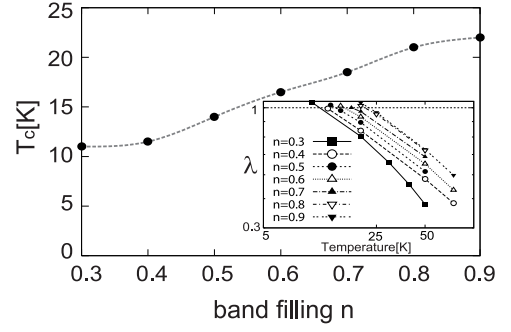}
 \caption{\label{fig3} The obtained $T_c$ for singlet pairing. The inset is $\lambda$ as functions of temperature. $U=2.0$ eV, $V_1=0$ are taken.}
\end{center}
\end{figure}

\begin{figure}
\begin{center}
 \includegraphics[width=6.8cm]{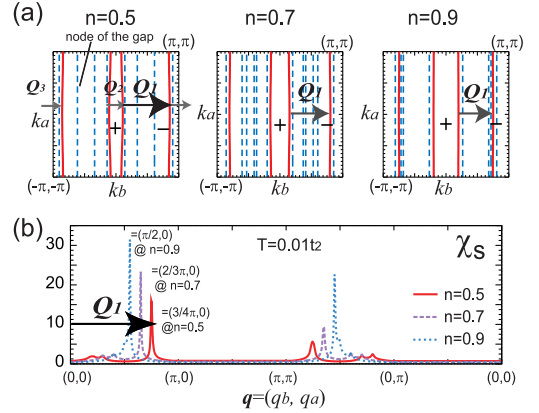}
\caption{\label{fig4}(Color online) (a) The FS and the nodes of the gap function, and (b) the spin susceptibility $\chi_s(\bvec{q})$ for $U=2.0$ eV and $V_1=0$ eV.}
\end{center}
\end{figure}

First, we present the result for $U=2.0$ eV without the off-site interaction.
The obtained $T_c$ is shown in Fig. \ref{fig3} as functions of $n$, and
the largest eigenvalue $\lambda$ vs. $T$ for various band fillings is
also shown in the inset. Finite values of $T_c$ are obtained in a wide range of
band filling, and its value gradually increases with increasing $n$. 
%The largest eigenvalue $\lambda$ for singlet pairing 
%is shown in Fig. \ref{fig3} as functions
%of $T$ for various band fillings, and the obtained band filling dependence of
%$T_c$ is also shown the inset.  Finite values of $T_c$
%are obtained in a wide range of band filling, and its value gradually 
%increases with increasing $n$. 
%if we assume the band filling of Pr247($\delta=0$) is around the $1/4$
%filling,\cite{Mizokawa, Takenaka} the finite $T_c$ doesn't agree with
%the experimental results. 
The FS (defined as $\varepsilon_k - \mu+{\rm Re}(\Sigma(k))=0$)
along with the nodes of the gap function, and the spin susceptibility,
$\chi_s(\bvec{q})$, are plotted in Fig. \ref{fig4}. Three kinds of 
nesting (hereafter specified by nesting vectors $\bvec{Q}_1$,$\bvec{Q}_2$ and
$\bvec{Q}_3$) cause the peaks of $\chi_s$. The most dominant nesting
$\bvec{Q}_1$ arises due to the nesting between the inner and the outer
sets of FS, and thus is a direct consequence of the presence of multiple
FS, while the other two originate from inner-inner or outer-outer FS
nesting. $\bvec{Q}_1$ changes from $\bvec{Q}_1=(3\pi/4,q_a)$ at $n=0.5$
to $(\pi/2,q_a)$ at $n=0.9$ and the values of peak itself increase with
increasing the band filling because of the FS nesting becomes
better. There are no nodes intersecting the FS, but the sign of the gap
changes its sign between the inner and the outer FS, which is a
consequence of the repulsive pairing interaction mediated by the spin
fluctuations at the nesting vector $\bvec{Q}_1$. Consequently, the
values of $\lambda$ and $T_c$  increases with the increase of the peak
value of $\chi_s$ at $\bvec{Q}_1$ upon increasing the band filling. We
will call this gap an extended $s$-wave hereafter.

The obtained $T_c$ of $\sim 20$ K agrees with the experimental
result, but the gradual band filling dependence of $T_c$ fails to
explain the ``switch on'' of the SC with oxygen reduction, i.e., the
increase of $n$ in the realistic band filling range. At least within the
present approach, the change of the topology of the inner FS (i.e.,
change of the number of Fermi points along the X-S line) does not
strongly affect SC.

\begin{figure}
\begin{center}
\includegraphics[width=6.4cm]{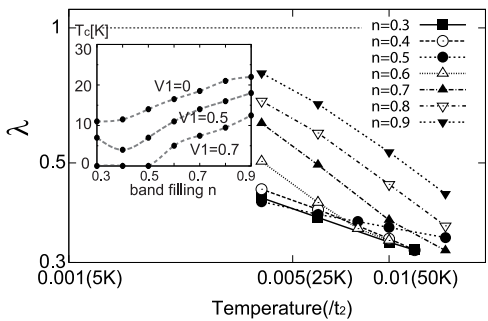}
\caption{\label{fig5}$\lambda$ vs. $T$ for $U=2.0$ eV, $V_1=0.7$
 eV. (Inset) The estimated values of $T_c$ obtained by linear extrapolation.}
\end{center}
\end{figure}

\begin{figure}
\begin{center}
\includegraphics[width=6.8cm]{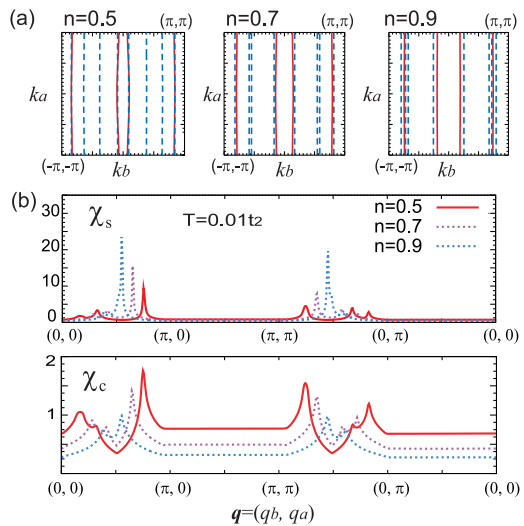}
\caption{\label{fig6}(Color online) (a) The FS and the nodes of the gap, 
and (b) the spin and the charge susceptibilities $\chi_s$ and 
$\chi_c$ for $U=2.0$ eV, $V_1=0.7$ eV.}
\end{center}
\end{figure}
\begin{figure}
\begin{center}
\includegraphics[scale=0.6]{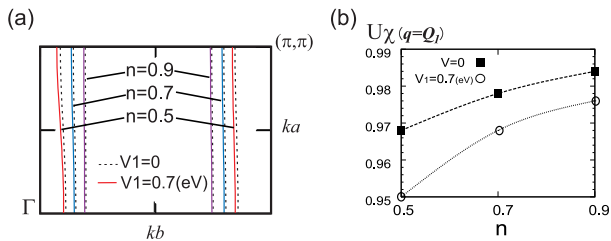}
\caption{\label{fig7} (Color online) (a) The FS and (b) the filling dependence of $U\chi(\bvec{Q}_1)$ for $V_1=0$ and $0.7$ eV. $U$ is fixed at 2.0 eV.}
\end{center}
\end{figure}

This result has motivated us to include the n.n. repulsion $V_1$.
Since the system size and the number of Matsubara frequencies are
limited, we have calculated $\lambda$ for $T>0.004t_2$ in order to assure the
convergence. Figure \ref{fig5} shows the obtained $\lambda$ as functions of
$T$ with $V_1=0.7$ eV. The value of $\lambda$ near $n=0.5$ is strongly 
suppressed compared to the result without $V_1$, and does not seem to
reach unity even at low temperatures, while for higher filling such a
tendency of suppression is smaller. %This result can be traced back to the
%deformation of the FS shown in %Fig.\ref{fig6}. nodes of the gap function, spin and charge susceptibilities. 
This result can be traced back to the deformation of the FS shown in
Fig.\ref{fig6}, where the nodes of the gap function and the spin and the charge
susceptibilities are shown. Although the positions of $\bvec{Q}_1$
and the gap function do not differ so much from the results for
$V_1=0$, the peak value of $\chi_s$, particularly for $n=0.5$, turns out
to be smaller. As seen in Fig. \ref{fig7} (a) and (b), where the
obtained FS and the values of $U\chi(\bvec{Q}_1)$ for $V_1=0$ and $0.7$
eV are plotted, this reduction of $\chi_s$ is caused by the degradation
of the nesting condition due to the renormalization of the band for $V_1
\neq 0$, which results in a stronger warping of the FS especially around
$1/4$ filling. The reduction of $\chi_s$ is reflected in the reduction
of SC especially near $n\simeq 0.5$.
%The positions of $\bvec{Q}_1$ and the pairing symmetries are not so
%change compared to the result for $V_1=0$. Note that the peak values of
%$\chi_s$, particularly for $n=0.5$, turn out to be lower than that
%obtained in $V_1=0$. This is the main reason for the suppression of
%$\lambda$ near the $1/4$ filling. In Fig \ref{fig7}(a) is the some
%elements of the uncorrelated fluctuation propagator matrix
%$\bar{\chi}(\bvec{q};ij)$, where $i (j)=1,2$,and 3 means $\Delta\bvec{R}$
%($\Delta\bvec{R}')=0,\hat{\bvec{x}}$, and $-\hat{\bvec{x}}$,
%respectively. The diagonal components $\chi(\bvec{q};11)\left(=\chi(\bvec{q};22)=\chi(\bvec{q};33)\right)$
%are the largest components and take high values at the nesting vectors
%$\bvec{q}=\bvec{Q}_{1,2,3}$. On the other hand,
%$(ij)=(12),(13),(21),(31)$ components are almost zero even at
%$\bvec{q}=\bvec{Q}_{1,2,3}$ because of the cancellation of phase factor
%at $\bvec{k} \in \bvec{k}_{FS}$ regardless the band fillings. If these
%components are regarded as zero, the maximum eigenvalues of
%$\bar{\chi_s}(\bvec{q};\Delta\bvec{R},\Delta\bvec{R}')$ is given as
%$\chi(\bvec{q};11)[1-U\chi(\bvec{q};11)]^{-1}$, so that $\chi_s(\bvec{q})$ is governed by $\chi(\bvec{q};11)$. The filling dependence of
%$U\chi(\bvec{q};11)$ at $\bvec{q}=\bvec{Q}_1$ as well as
%$\chi(\bvec{q})$ obtained in $V_1=0$ are plotted in
%Fig. \ref{fig7}(b). Note that, at $n \simeq 0.5$, the lowering of
%$\chi(\bvec{Q}_1;11)$, which leads the reduction of
%$\chi_s(\bvec{Q}_1)$, is caused by the nesting conditions being worse
%due to the warp of the renormalized FS as seen in Fig.\ref{fig7}(c). 
As for the {\it direct} contribution of the charge fluctuation to the 
pairing interaction through the $\frac{1}{2}\bar{V}_d\bar{\chi}_c\bar{V}_d$ 
term, although $\chi_c(\bvec{Q}_1)$ near 
$1/4$ filling is relatively large compared to those for other band fillings,
the peak value itself is much smaller than that of the spin susceptibility,
so that this contribution should hardly affect SC.
Thus the main contribution of the charge fluctuation is through the 
self-energy renormalization.

Although we cannot obtain $T_c$ within the investigated temperature
range, judging the obtained temperature dependence of $\lambda$, it
does not seem unnatural to linearly extrapolate $\log(\lambda)$ 
up to $\lambda=1$ against $\log(T)$ 
in order to estimate the value of $T_c$. In the inset of Fig. \ref{fig5}, the
estimated values of $T_c$ are shown. If we assume that the band filling
of Pr247/$\delta=0$ is $n \simeq 0.5$, this result shows qualitative
agreement with the experimental result in that the SC appears with
moderate electron doping. 

\begin{figure}
\begin{center}
\includegraphics[width=4.6cm]{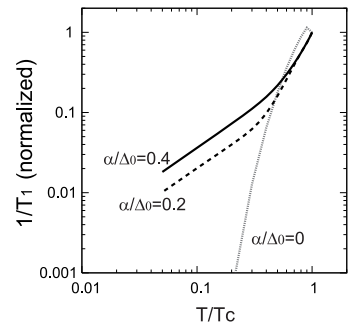}
\caption{\label{fig9}The calculated $1/T_1$ for several values of pair
 breaking parameter for $n=0.7$, $U=2.0$ eV, $V_1=0.7$ eV. 
}
\end{center}
\end{figure}

Finally, we discuss the validity of the obtained form of the gap
function in the light of the experimental results. 
The NQR measurement has observed $1/T_1 \propto T^2$ 
($T_1$ is the spin-lattice relaxation rate) without
exhibiting a coherence peak below $T_c$.\cite{Sasaki} This result seems to
suggest that the nodes of the gap intersect the FS. 
However, if we assume that a 
small amount of oxygen defects controlling the band filling work as
scatterers, we find that the observed $1/T_1$ can be accounted for as follows. 
Taking into account the 
effect of the impurities and the defects in the unitarity limit, 
we have calculated
$1/T_1$ for the gap functions and the FS obtained by FLEX.\cite{Hotta}
As for the temperature dependence of the gap function,
$\Delta(\bvec{k},T)=\Delta_0\phi(\bvec{k})\tanh(2\sqrt{(T_c/T)-1})$ 
with $\Delta_0/(k_BT_c)=4$ is assumed, where  
$\phi(\bvec{k})$ is proportional 
to the gap function obtained within the FLEX+Eliashberg eq. approach.
In Fig. \ref{fig9}, the obtained $1/T_1$
is shown for $U=2.0$ eV, $V_1=0.7$ eV. The values of $\alpha/\Delta_0$
are chosen as $0, 0.2$ and $0.4$, where $\alpha=c/\pi N_0(0)$ is the pair
breaking parameter, $c$ is the impurity concentration and $N_0(0)$ is
the density of states at the Fermi level in the normal state. When
$\alpha/\Delta_0=0$, $1/T_1$ decays exponentially since a full gap opens
on the FS. For finite $\alpha/\Delta_0$, $1/T_1$ shows a power-law-like
decay without exhibiting a coherence peak, which resembles the
experimental result. Therefore, from these results, we can safely say
that the observed $1/T_1$ does not necessary rule out the possibility of
the fully gapped extended s-wave pairing.  
% Of course, in the situation like $n=0.5$ of Fig. \ref{fig6}(a), $1/T_1$
%decays as $T^2$ as shown in the figure, but in this case, finite $T_c$
%is not obtained.

To summarize, we have investigated the superconductivity of
Pr$_2$Ba$_2$Cu$_4$O$_{15-\delta}$ using an extended Hubbard model on 
a Q1D double chain with the TB parameters determined from the LDA
calculation. %From the discussion on the band structure, it is found that
%the change in the number of Fermi points occurs only in the particular
%direction (X-S) and there is slightly two dimensional nature. 
By applying the FLEX to the model and solving the Eliashberg equation,
in the absence of $V_1$, we have obtained singlet superconductivity with
moderately varying $T_c$ in a wide band filling range. 
%By introducing the realistic values of n.n and n.n.n interaction $V_1$
%and $V_2$, 
By introducing the n.n interaction $V_1$, the superconductivity is
suppressed at a band filling range near $1/4$-filling, 
which at least qualitatively explains the $\delta$ 
dependence of superconductivity.
%%
%%The vanishing of the superconductivity at $\delta > 0.6$ may related to
%%the Mott insulating state. In this study, the ambiguity of the band
%%dispersion, and thus, the band fillings still remain. Further
%%quantitative study will be needed in this point. 
NQR relaxation rate $1/T_1$ is also calculated taking into account the
effect of the impurities and the defects. 
Power-law-like decaying $1/T_1$ resembling
the experimental result is obtained with a moderate impurity concentration.

We should comment on the higher $\delta$ regime ($\delta > 0.5$),
where $T_c$ is found to 
decrease with increasing $\delta$ and vanishes 
at $\delta \sim 0.6$.\cite{Matsukawa} 
Considering that the nesting vector $\bvec{Q}_1$,
$\bvec{Q}_2$ and $\bvec{Q}_3$ all become close to $(\pi/2,q_a)$ near half
filling (see Fig. \ref{fig4} and Fig. \ref{fig6}), umklapp processes 
can be allowed to result in a Mott insulating state,\cite{Fabrizio} 
which cannot be dealt with in the present method. More
study on this issue will be carried out elsewhere.
In a zigzag system as in the present case, 
next n.n. interactions may also play some role as suggested in
experimental and theoretical studies. \cite{Fujiyama, Seo}
Unfortunately, the calculation including such interactions within the
present approach has to be restricted to small system sizes. Therefore
the study on this effect also remains as future study.

Present results somewhat resemble that of the recent study on
Na$_x$CoO$_2\cdot y$H$_2$O in that the extended $s$-wave
superconductivity occurs due to the inner-outer FS nesting in
a disconnected FS system.\cite{KN} Further studies on such systems 
may open interesting new physics.

The authors thank N. Yamada for his encouragement throughout this study, T. 
Oguchi for discussions on the band structure, and T. Nishida for useful
suggestions. We also acknowledge illuminating discussions with Y. \={O}no,
K. Sano, T. Habaguchi, Y. Yamada, and T. Mizokawa. Numerical calculations were
performed at the facilities of Supercomputer Center, ISSP, University of
Tokyo. This work was supported by Grants-in-Aid for Scientific Research
from the Ministry of Education, Culture, Sports, Science and Technology
of Japan.


\begin{thebibliography}{99}
\bibitem{comment} 
Although the band structure in refs.\onlinecite{Ambrosch} and
	\onlinecite{Yu} is that for Y124, a very similar band structure
	has been recently obtained for Pr247 by T. Habaguchi and Y. \={O}no, private communications.
\end{thebibliography}
\end{document}